\documentstyle[11pt]{article}

  \let\n=\nu

\def\nn{\nonumber} \def\bd{\begin{document}} \def\ed{\end{document}}
\def\ds{\documentstyle} \let\fr=\frac \let\bl=\bigl \let\br=\bigr
\let\Br=\Bigr \let\Bl=\Bigl 
\let\bm=\bibitem
\let\na=\nabla
\let\pa=\partial \let\ov=\overline 
\newcommand{\be}{\begin{equation}} 
\newcommand{\ee}{\end{equation}} 
\def\ba{\begin{array}}
\def\ea{\end{array}}
\def\ft#1#2{{\textstyle{{\scriptstyle #1}\over {\scriptstyle #2}}}}
\def\fft#1#2{{#1 \over #2}}
\def\del{\partial}
\def\vp{\varphi}
\def\sst#1{{\scriptscriptstyle #1}}
\def\oneone{\rlap 1\mkern4mu{\rm l}}
\def\td{\tilde}
\def\wtd{\widetilde}
\def\ie{\rm i.e.\ }
\def\dalemb#1#2{{\vbox{\hrule height .#2pt
        \hbox{\vrule width.#2pt height#1pt \kern#1pt
                \vrule width.#2pt}
        \hrule height.#2pt}}}
\def\square{\mathord{\dalemb{6.8}{7}\hbox{\hskip1pt}}}
\newcommand{\ho}[1]{$\, ^{#1}$}
\newcommand{\hoch}[1]{$\, ^{#1}$}
\newcommand{\bea}{\begin{eqnarray}} 
\newcommand{\eea}{\end{eqnarray}} 
\newcommand{\ra}{\rightarrow}
\newcommand{\lra}{\longrightarrow}
\newcommand{\Lra}{\Leftrightarrow}
\newcommand{\ap}{\alpha^\prime}
\newcommand{\bp}{\tilde \beta^\prime}
\newcommand{\tr}{{\rm tr} }
\newcommand{\Tr}{{\rm Tr} } 
\def\0{{\sst{(0)}}}
\def\1{{\sst{(1)}}}
\def\2{{\sst{(2)}}}
\def\3{{\sst{(3)}}}
\def\4{{\sst{(4)}}}
\def\5{{\sst{(5)}}}
\def\6{{\sst{(6)}}}
\def\7{{\sst{(7)}}}
\def\8{{\sst{(8)}}}
\def\n{{\sst{(n)}}}
\def\tV{\widetilde V}
\def\tW{\widetilde W}
\def\tH{\widetilde H}
\def\tE{\widetilde E}
\def\tF{\widetilde F}
\def\tA{\widetilde A}
\def\im{{{\rm i}}}
\def\tY{{{\wtd Y}}}
\def\ep{{\epsilon}}
\def\vep{{\varepsilon}}
\def\R{\rlap{\rm I}\mkern3mu{\rm R}}

\newcommand{\NP}{Nucl. Phys. }
\newcommand{\tamphys}{\it Center for Theoretical Physics,
Texas A\&M University, College Station, TX 77843}
\newcommand{\upenn}{\it Dept. of Phys. and Astro.,
University of Pennsylvania,
Philadelphia, PA 19104}

\newcommand{\auth}{M. Cveti\v{c}\hoch{\dagger1}, H. L\"u\hoch{\dagger1},
C.N. Pope\hoch{\ddagger2}, A. Sadrzadeh\hoch{\ddagger} and 
  T.A. Tran\hoch{\ddagger}}

\begin{document}
\begin{flushright}
\hfill{CTP TAMU-08/00 \\
UPR/879-T \\
March 2000}\\
\hfill{\bf hep-th/0003103}\\
\end{flushright}

\vspace{10pt}

\begin{center}
{\large {\bf Consistent $SO(6)$ Reduction Of Type IIB Supergravity on $S^5$}}

\vspace{20pt}

\auth

\vspace{10pt}
{\hoch{\dagger}\upenn}

\vspace{10pt}
{\hoch{\ddagger}\tamphys}

\vspace{30pt}

\underline{ABSTRACT}
\end{center}

         Type IIB supergravity can be consistently truncated to the
metric and the self-dual 5-form.  We obtain the complete non-linear
Kaluza-Klein $S^5$ reduction Ansatz for this theory, giving rise to
gravity coupled to the fifteen Yang-Mills gauge fields of $SO(6)$ and
the twenty scalars of the coset $SL(6,\R)/SO(6)$.  This provides a
consistent embedding of this subsector of $N=8$, $D=5$ gauged
supergravity in type IIB in $D=10$.  We demonstrate that the
self-duality of the 5-form plays a crucial role in the consistency of
the reduction.  We also discuss certain necessary conditions for a
theory of gravity and an antisymmetric tensor in an arbitrary
dimension $D$ to admit a consistent sphere reduction, keeping all the
massless fields.  We find that it is only possible for $D=11$, with a
4-form field, and $D=10$, with a 5-form.  Furthermore, in $D=11$ the
full bosonic structure of eleven-dimensional supergravity is required,
while in $D=10$ the 5-form must be self-dual.  It is remarkable that
just from the consistency requirement alone one would discover $D=11$ and
type IIB supergravities, and that $D=11$ is an upper bound on the
dimension.

{\vfill\leftline{}\vfill
\vskip 10pt \footnoterule {\footnotesize \hoch{1} Research supported
in part by DOE grant DOE-FG02-95ER40893
\vskip  -12pt} \vskip   14pt
{\footnotesize
        \hoch{2}        Research supported in part by DOE
grant DOE-FG03-95ER40917 \vskip -12pt}  \vskip  14pt
}

\pagebreak
\setcounter{page}{1}

\section{Introduction}

     Non-trivial Kaluza-Klein sphere reductions of supergravity
theories have been studied in a number of contexts.  Long ago it was
demonstrated \cite{dwn} that the linearised analysis of the zero-mode
fluctuations of the $S^7$ reduction of $D=11$ supergravity
\cite{dp,DNP} can be extended to a fully non-linear and consistent
embedding of maximal $D=4$ gauged $SO(8)$ supergravity in $D=11$.  The
fact that the truncation to the zero-mode sector is consistent,
despite the non-linearities of the theory, is rather non-trivial since
there appears to be no group-theoretic understanding of why it should
work.

   More recently, a similar, although more explicit, demonstration of
the consistency of the $S^4$ reduction from $D=11$ to the maximal
$D=7$ gauged $SO(5)$ supergravity was given \cite{vann1,vann2}.  The
consistent reduction of massive type IIA supergravity on a locally
$S^4$ space, to give the $N=2$ gauged $SU(2)$ supergravity in $D=6$, has
also been obtained \cite{d6gauge}.  In addition, certain reductions to
truncations of the maximal gauged supergravities in various dimensions
have also been constructed.  These have the advantage of being
considerably simpler than the maximal theories, allowing the reduction
Ansatz to be presented in a more explicit form.  Cases that have been
constructed include the $N=2$ gauged $SU(2)$ supergravity in $D=7$
\cite{d7gauge}; the $N=4$ gauged $SU(2)\times U(1)$ supergravity in
$D=5$ \cite{d5gauge}; the $N=4$ gauged $SO(4)$ supergravity in $D=4$
\cite{d4gauge}; and maximal abelian truncations in $D=4$, 5 and 7
\cite{ten}.  One can also consider non-supersymmetric truncations of
the maximal gauged supergravities.  In \cite{cglp,clps} the consistent
truncations of the $D=7$, $D=5$ and $D=4$ supergravities to the
graviton plus scalar subsectors comprising only the diagonal scalars
in the $SL(N,\R)/SO(N)$ scalar submanifolds were considered (with
$N=5$, 6 and 8 respectively), and the consistent embeddings in $D=11$
and $D=10$ were constructed.

   Although the full consistency of the $S^7$ and $S^4$ reductions of
$D=11$ supergravity has essentially been demonstrated, no similar
complete result exists for the $S^5$ reduction of type IIB
supergravity.  The field content of the full $N=8$ gauged supergravity
\cite{grw} consists of gravity; fifteen $SO(6)$ gauge fields; twelve
2-form gauge potentials in the $6$ and $\overline{6}$ representations of
$SO(6)$; 42 scalars in the $1+1+20' + 10 +\overline{10}$ representations of
$SO(6)$, and the fermionic superpartners.  It is believed that this
can arise from an $S^5$ reduction of type IIB supergravity; at the
linearised level, the reduction Ansatz was given in \cite{krn}.
However, at the full non-linear level, the only complete
demonstrations so far are for the consistent embedding of the maximal
abelian $U(1)^3$ truncation \cite{ten}, the $N=4$ gauged $SU(2)\times
U(1)$ truncation \cite{d5gauge}, and the scalar truncation in
\cite{cglp,clps}.  The full metric Ansatz was conjectured in
\cite{kpw}.

   In this paper, we obtain the consistent reduction Ansatz for a
different truncation of the maximal gauged $D=5$ supergravity.  One
can consistently set the $1+1+10 +\overline{10}$ of scalars to zero, at the
same time as setting the $6+\overline{6}$ of 2-form potentials to zero.
The bosons that remain, namely gravity, the $15$ Yang-Mills fields,
and the $20'$ of scalars, come just from the metric plus the self-dual
5-form sector of the original type IIB theory.  We shall obtain
complete results for the consistent embedding of this subsector of the
gauged $N=8$ theory, with all fifteen of the $SO(6)$ gauge fields
$A_\1^{ij}$ , and the twenty scalars that parameterise the full
$SL(6,\R)/SO(6)$ submanifold of the complete scalar coset.  These can
be parameterised by a unimodular symmetric tensor $T_{ij}$.

    Another way of expressing the truncation that we shall consider in
this paper is as follows.  The type IIB theory itself can be
consistently truncated in $D=10$ so that just gravity and the
self-dual 5-form remain.  The fifteen Yang-Mills gauge fields and
twenty scalars that we retain in our Ansatz are the full set of
massless fields associated with Kaluza-Klein reduction from this
ten-dimensional starting point.  (The counting of massless fields is
the same as the one that arises from a toroidal reduction from the
same ten-dimensional starting point.)  We shall see below that the
self-duality condition on the 5-form plays an essential r\^ole in the
consistency of the $S^5$ reduction.

   This subsector of type IIB supergravity is particularly relevant
for the AdS/CFT correspondence \cite{malda,gkp,wit}, because it
is the metric and the self-dual 5-form that couple to the D3-brane.  

   We also address the more general question of the circumstances
under which a theory admits a consistent sphere reduction.  We show
that in a $D$-dimensional theory of gravity coupled to an $n$-form
field strength, a consistent $S^n$ reduction that retains the full set
of $SO(n+1)$ Yang-Mills fields together with coupled massless scalars
is possible only when $D=11$ and $n=4$ or $7$, or in $D=10$ with
$n=5$.  Furthermore, in $D=11$ the theory has to be the bosonic sector
of eleven-dimensional supergravity, with the $FFA$ term, while in
$D=10$ the 5-form must be self-dual.  In all three cases the full set
of massless scalars includes a subset $T_{ij}$ described by the coset
$SL(n+1,\R)/SO(n+1)$.  (Such a coset structure is absent for any other
values of $(D,n)$, and so it would not be appropriate to look for a
consistent reduction Ansatz with scalars $T_{ij}$ of
$SL(n+1,\R)/SO(n+1)$ for generic $(D,n)$.)  Of the three cases where
such a consistent sphere reduction is possible, the ten-dimensional
one is singled out as the only case where the consistent reduction
includes only gravity plus the $SO(n+1)$ gauge fields $A_\1^{ij}$ and
the scalars $T_{ij}$.  By contrast, for $D=11$ reduced on $S^4$ one
must additionally retain a set of five 2-form potentials, while for
$D=11$ reduced on $S^7$ one must instead additionally retain 35
pseudoscalars as well as the 35 scalars $T_{ij}$, in order to achieve
consistent reductions.\footnote{This is related to the fact that 
one can consistently truncate five-dimensional maximal gauged
supergravity to the $SO(6)$ gauge fields plus scalars of the coset
$SL(6,\R)/SO(6)$.  By contrast, the analogous truncations cannot be
performed in the maximal gauged supergravities in $D=7$ and $D=4$.}

\section{The $SO(6)$ reduction Ansatz on $S^5$}

   We parameterise the fields for this truncated theory as follows.
The twenty scalars, which are in the $20'$ representation of $SO(6)$,
are represented by the symmetric unimodular tensor $T_{ij}$, where $i$
is a 6 of $SO(6)$.  The fifteen $SO(6)$ Yang-Mills gauge fields will
be represented by the 1-form potentials $A_\1^{ij}$, antisymmetric in
$i$ and $j$.  The inverse of the scalar matrix $T_{ij}$ is denoted by
$T^{-1}_{ij}$.  In terms of these quantities, we find that the
Kaluza-Klein reduction Ansatz is given by
\bea
d\hat s_{10}^2 &=& \Delta^{1/2}\, ds_5^2 + g^{-2}\, 
\Delta^{-1/2}\, T^{-1}_{ij}\, 
  D\mu^i\, D\mu^j\,,\label{metans}\\
\hat H_\5 &=& \hat G_\5 + {{\hat *}\hat G_\5}\,,\label{hans}\\
\hat G_\5 &=& -g\, U\, \ep_\5 + g^{-1}\, (T^{-1}_{ij}\, {*D}\, T_{jk})\wedge 
(\mu^k\, D\mu^i)\nn\\
&& -\ft12 g^{-2}\, 
T^{-1}_{ik}\, T^{-1}_{j\ell}\, {*F_\2}^{ij}\wedge
D\mu^k\wedge D\mu^\ell\,,\label{gans}\\
{{\hat *}\hat G_\5} &=& \fft1{5!}\, \ep_{i_1\cdots i_6}\, \Big[
g^{-4}\, U\, \Delta^{-2}\, D\mu^{i_1}\wedge \cdots \wedge D\mu^{i_5}\,
\mu^{i_6}\nn\\
&& -5 g^{-4}\, \Delta^{-2}\, D\mu^{i_1}\wedge \cdots \wedge D\mu^{i_4}
\wedge DT_{i_5 j}\, T_{i_6 k}\, \mu^j\, \mu^k \nn\\
&&- 10 g^{-3}\, \Delta^{-1}\, 
 F_\2^{i_1 i_2}\wedge D\mu^{i_3}\wedge D\mu^{i_4}\wedge D\mu^{i_5}\, 
T_{i_6 j}\, \mu^j \Big]\,,\label{gdualans}
\eea
where
\bea
&&U \equiv 2 T_{ij}\, T_{jk}\, \mu^i\, \mu^k -\Delta\, T_{ii}\,, \qquad 
\Delta \equiv T_{ij}\, \mu^i\, \mu^j\,,\nn\\
&&F_\2^{ij} = dA_\1^{ij} + g\, A_\1^{ik}\wedge A_\1^{kj}\,,
\qquad DT_{ij} \equiv dT_{ij} + g\, A_\1^{ik}\, T_{kj} + g\, A_\1^{jk}\, 
T_{ik}\,,\nn\\
&& \mu^i\, \mu^i = 1\,,\qquad 
D\mu^i \equiv d\mu^i +g\,  A_\1^{ij}\, \mu^j\,,
\eea
and $\ep_\5$ is the volume form on the five-dimensional spacetime.
Note that ${{\hat *}\hat G_\5}$ is derivable from the given
expressions (\ref{metans}) and (\ref{gans}); we have presented it here
because it is quite an involved computation.  The coordinates $\mu^i$,
subject to the constraint $\mu^i\, \mu^i=1$, parameterise points in
the internal 5-sphere.  In obtaining the above Ansatz we have been
guided by previous results in the literature, including the $S^4$
reduction Ansatz from $D=11$ that was constructed in \cite{vann1,vann2}.

   It is consistent to truncate the fields of type IIB supergravity to
the metric and self-dual 5-form $\hat H_\5$.  The ten-dimensional
equations for motion for these fields are then given by
\bea
\hat R_{MN} &=& \fft1{96}\, \hat H_{MPQRS}\, \hat H_N{}^{PQRS}\,,\nn\\
d\hat H_\5 &=& 0\,.
\eea
The Ansatz presented above satisfies these equations of motion if and
only if the five-dimensional fields satisfy the equations
\bea
D(T^{-1}_{ik}\, T^{-1}_{j\ell}\, {*F_\2^{k\ell})} &=&
-2g\, T^{-1}_{k[i}\, {*D}T_{j]k} - \ft18 \ep_{ij k_1\cdots k_4}\, 
F_\2^{k_1 k_2}\wedge F_\2^{k_3 k_4}\,,\nn\\
D(T^{-1}_{ik}\, {*D} T_{kj}) &=& -2g^2\, (2T_{ik}\, T_{jk} - T_{ij}\,
T_{kk})\,\ep_\5 + T^{-1}_{ik}\, T^{-1}_{\ell m}\, {*F_\2^{\ell k}}
\wedge F_\2^{mj}\label{d5eom}\\
&&-\ft16 \delta_{ij}\, \Big[ -2g^2\, (2T_{k\ell}\, T_{k\ell} -
(T_{kk})^2)\, \ep_\5 +
T^{-1}_{pk}\, T^{-1}_{\ell m}\, {*F_\2^{\ell k}} \wedge
F_\2^{mp} \Big]\,,\nn
\eea
together with the five-dimensional Einstein equation.  These equations
of motion can all be derived from the five-dimensional Lagrangian
\bea
{\cal L}_5 &=& R\, {*\oneone} - \ft14 T^{-1}_{ij}\, {*D T_{jk}}\wedge
T^{-1}_{k\ell}\, DT_{\ell i} - \ft14 T^{-1}_{ik}\,
T^{-1}_{j\ell}\, {* F_\2^{ij}}\wedge F_\2^{k\ell} 
-V\, {*\oneone}\label{d5lag}\\
&& \!\!\! - \ft1{48}\, \ep_{i_1\cdots i_6}\, 
\Big(F_\2^{i_1 i_2}\, F_\2^{i_3 i_4}\, A_\1^{i_5 i_6} -
 g\, F_\2^{i_1 i_2}\, A_\1^{i_3 i_4}\, A_\1^{i_5 j}\, A_\1^{j i_6} 
+\ft25 g^2\, A_\1^{i_1 i_2} \, A_\1^{i_3 j}\, A_\1^{j i_4}\, A_\1^{i_5
k}\, A_\1^{k i_6} \Big)\,,\nn
\eea
where the potential $V$ is given by
\be
V = \ft12 g^2\, \Big(2 T_{ij}\, T_{ij} - (T_{ii})^2 \Big)\,.
\ee
In (\ref{d5lag}) we have omitted the wedge symbols in the final
topological term, to economise on space.  The Lagrangian is in agreement 
with the one for five-dimensional gauged $SO(6)$ supergravity in
\cite{grw}.

   To see this, we look first at the ten-dimensional equation $d{\hat
H_\5}=0$, with the Ansatz given above.  The terms involving structures
of the form $X^{ij}_\4 \wedge D\mu^i\wedge D\mu^j$, where $X^{ij}_\4$
represents a 4-form in the five-dimensional spacetime, give rise to
the Yang-Mills equations above.  Contributions of this kind come from
$d{\hat G_\5}$ and also from the final term in $d{\hat *\hat G_\5}$.
The terms involving structures of the form $X^{ij}_\5\wedge (\mu^i\,
D\mu^j)$, where $X_\5^{ij}$ represents a 5-form in the five-dimensional
spacetime, give rise to the scalar equations of motion in
(\ref{d5eom}).  Since $\mu^i\, D\mu^i = \mu^i\, d\mu^i = \ft12
d(\mu^i\, \mu^i)=0$, there is a trace subtraction in the scalar
equation, and we read off $X_\5^{ij} - \ft16 \delta_{ij} X_\5^{kk}=0$
as the five-dimensional equation of motion.  Contributions of this
structure come only from $d\hat G_\5$.  Finally, all other structures
arising from calculating $d{\hat H_5}=0$ vanish identically, without
the use of any five-dimensional equations of motion.  In deriving
these results one needs to make extensive use of the Schoutens
over-antisymmetrisation identity, $\ep_{[i_1\cdots i_6}\, V_{i_7]}=0$.

   It is worth remarking that it is essential for the consistency of
the reduction Ansatz that the 5-form $\hat H_\5$ should be self-dual.
One cannot simply consider a reduction Ansatz for a ten-dimensional
theory consisting of gravity plus a non-self-dual 5-form, whose Ansatz
is given by $\hat G_\5$ in (\ref{gans}).  Although the Bianchi
identity $d\hat G_\5=0$ would give perfectly acceptable equations of
motion for $F_\2^{ij}$ and $T_{ij}$, the field equation $d{\hat *\hat
G_\5}=0$ would produce the (unacceptable) constraint
\be
\ep_{ijk_1\cdots k_4}\, F_\2^{k_1 k_2}\wedge F_\2^{k_3 k_4}=0\,.
\label{ffterm}
\ee
It is only by combining $\hat G_\5$ and ${\hat *\hat G_\5}$ together
into the self-dual field $\hat H_\5$ that a consistent
five-dimensional result is obtained, with (\ref{ffterm}) now combining
with terms from $d\hat G_\5$ to form part of the five-dimensional
Yang-Mills equations given in (\ref{d5eom}).\footnote{If one were to
consider the reduction of a ten-dimensional theory with a
non-self-dual 5-form there would be additional fields present in a
complete massless truncation.  These would comprise $10=4+6$ vector
potentials and $5=1+4$ scalars.  However the inclusion of these fields
would still not achieve a consistent reduction Ansatz, since the current
(\ref{ffterm}), which is in the $15$ of $SO(6)$, could still not
acquire an interpretation as a source term for the additional fields.
Thus the additional requirement of self-duality seems to be essential
for consistency.  (Of course anti-self-duality would be equally good.)}
It is interesting, therefore, that self-duality of the 5-form is
apparently forced on us by the requirements of Kaluza-Klein
consistency, if we try to ``invent'' a ten-dimensional theory that can
be reduced on $S^5$.  Thus once again we see that supersymmetry and
Kaluza-Klein consistency for sphere reductions seem to go hand in
hand.\footnote{Of course the initial truncation of the type IIB theory
to its gravity plus self-dual 5-form sector is itself a
non-supersymmetric one, but the crucial point is that the consistency
of the Kaluza-Klein $S^5$ reduction is singling out a starting point
that is itself a subsector of a {\it supersymmetric} theory.}

   The Kaluza-Klein $S^5$ reduction that we have obtained here retains
the full set of massless fields that can result from the reduction of
gravity plus a self-dual 5-form in $D=10$.  In other words, after the
initial truncation of the type IIB theory in $D=10$, no further
truncation of massless fields has been performed.

   It should also be emphasised that it would be inconsistent to omit
the fifteen $SO(6)$ gauge fields when considering the embedding of the
twenty scalars $T_{ij}$.  This can be seen from the Yang-Mills
equations in (\ref{d5eom}), which have a source term $g\,
T^{-1}_{k[i}\, {*D}T_{j]k}$ appearing on the right-hand side.  This is
a quite different situation from a toroidal reduction, where it is
always consistent to truncate to the scalar sector, setting the gauge
fields to zero.  The new feature here in the sphere reduction is that
the scalar fields are charged under the gauge group.  This is a
general feature of all the sphere reductions, to $D=4$, $D=5$ and
$D=7$, and thus in all cases it is inconsistent to include the full
set of scalar fields without including the gauge fields as well (see also
\cite{vann1,vann2,nv}).  (One {\it can} consistently truncate to the
diagonal scalars in $T_{ij}$, setting all the gauge fields to zero, as
in \cite{cglp,clps}, since then $T^{-1}_{k[i}\, {*D}T_{j]k}=0$.)

    Our testing of the consistency of the reduction Ansatz
(\ref{metans})-(\ref{gdualans}) has so far been restricted to
checking the equations of motion for $\hat H_\5$.  A full testing of
the consistency of the ten-dimensional Einstein equations would be
quite involved, and will be addressed in future work.\footnote{The
experience in all previous work on consistent reductions is that
although the actual checking of the higher-dimensional Einstein
equations is the most difficult part from a computational point of
view, the consistency seems to be assured once it has been achieved
for the equations of motion for the antisymmetric tensor fields, which
itself is an extremely stringent requirement.}  In the next section we
shall show that the reduction Ansatz that we have obtained in this
paper reduces, with appropriate additional truncations, to results
that have been obtained previously.  Since the complete consistency
was proven in these earlier results, including the ten-dimensional
Einstein equations, this provides further supporting evidence for the
complete consistency of the Ansatz that we have constructed here.

   Finally in this section, we remark that the $S^5$ reduction that we
have constructed here is also
consistent if we include the dilaton $\hat\phi$ and axion $\hat\chi$ 
of the type IIB theory.  These simply reduce according to the Ans\"atze
\be
\hat\phi = \phi\,,\qquad \hat \chi= \chi\,,
\ee
where the unhatted quantities denote the fields in five dimensions.  In this
reduction they do not appear in the previous Ans\"atze for the metric and
self-dual 5-form, and in $D=5$ they just give rise to the additional
$SL(2,\R)$-invariant Lagrangian
\be
{\cal L}_{(\phi,\chi)} = -\ft12 {*d\phi}\wedge d\phi - \ft12 e^{2\phi}\, 
{*d\chi}\wedge d\chi\,,
\ee
which is added to (\ref{d5lag}).  Note in particular that $\phi$ and $\chi$
do not appear in the five-dimensional scalar potential $V$.

\section{Truncations to previous results}

   We can consider three different truncations of the reduction scheme
of the previous section, in order to make contact with previous
results in the literature.  The first of the three involves truncating
the twenty scalars $T_{ij}$ of the $SL(6,\R)/SO(6)$ coset to the
diagonal subset
\be
T_{ij} = {\rm diag}\, (X_1, X_2, X_3, X_4, X_5, X_6)\,,
\ee
where $\prod_i X_i=1$, and setting all the fifteen gauge fields
$F_\2^{ij}$ to zero.  This reduces to the embedding that was obtained
in \cite{cglp}, for which the complete proof of consistency was
constructed in \cite{clps}.  

   The second possible truncation involves reducing the scalar sector
still further, to
\be
T_{ij} = {\rm diag}\, (\wtd X_1, \wtd X_1, \wtd X_2, \wtd X_2, 
\wtd X_3, \wtd X_3)\,,
\ee
where $\prod_a \wtd X_a=1$, but now retaining the three $U(1)$ gauge fields
$F_\2^{12}$, $F_\2^{34}$ and $F_\2^{56}$ of the maximal abelian
$U(1)^3$ subgroup of $SO(6)$.  (It is easy to see, by looking at the
five-dimensional equations of motion in (\ref{d5eom}), that this is a
consistent truncation.)  The truncated theory is supersymmetric, and
describes five-dimensional $U(1)$-gauged simple supergravity coupled to two
$U(1)$ vector multiplets.  The consistent embedding of this theory in
type IIB supergravity was obtained in \cite{ten}.

   The third possible truncation involves retaining just a single
scalar field $X$, by taking
\be
T_{ij} = {\rm diag}\, (X,X,X,X, X^{-2}, X^{-2})\,,\label{xscalar}
\ee
at the same time retaining only the gauge fields of $SU(2)\times
U(1)$.  It is convenient now to take the $SO(6)$ indices $i$ to range
over $0\le i\le 5$.  We take all the gauge potentials $A_\1^{ij}$ to
be zero except for the following:
\bea
&&A_\1^{01}= A_\1^{23}= \ft1{\sqrt2}\, A_\1^1\,,\quad
A_\1^{02}= A_\1^{31}= -\ft1{\sqrt2}\, A_\1^2\,,\quad
A_\1^{03}= A_\1^{12}= \ft1{\sqrt2}\, A_\1^3\,,\nn\\
&&A_\1^{45} = B_\1\,.\label{gaugefields}
\eea
We also parameterise the coordinates $\mu^i$ of the internal 5-sphere
as follows:
\bea
&&\mu^0 + \im\, \mu^3 = \cos\xi\, \cos\ft12\theta\, e^{\im\,
(\psi+\phi)/2}\,,\qquad
\mu^1+ \im\, \mu^2 = \cos\xi\, \sin\ft12\theta\, e^{\im\,
(\psi-\phi)/2}\,,\nn\\
&&\mu^4+\im\, \mu^5 = \sin\xi\, e^{\im\, \tau}\,.\label{mus}
\eea

    Substituting (\ref{xscalar}), (\ref{gaugefields}) and (\ref{mus})
into the metric Ansatz (\ref{metans}), we obtain
\bea
d\hat s^2_{10} &=&
\Delta^{1/2}\, ds_5^2 +  g^{-2} \, X\,
\Delta^{1/2}\, d\xi^2 +
g^{-2}\Delta^{-1/2}\, X^2\, s^2\, \Big(d\tau - g\, B_\1\Big)^2 \nn\\
& &+ \ft14 g^{-2}\, \Delta^{-1/2}\, X^{-1}\,
c^2\, \sum_i (\sigma^i - \sqrt{2}g\, A_\1^i)^2\,,
\eea
where $c\equiv \cos\xi$, $s\equiv \sin\xi$, $\Delta= X^{-2}\, s^2 +
X\, c^2$,  $h^i\equiv \sigma^i - \sqrt{2} g\, A_\1^i$, and the $\sigma_i$ 
denote the three left-invariant 1-forms of $SU(2)$, given by 
$\sigma_1+\im\, \sigma_2 = e^{-\im\psi}\,
(d\theta + \im\, \sin\theta\, d\phi)$, $\sigma_3=d\psi +
\cos\theta\, d\phi$. This is the metric Ansatz obtained in
\cite{d5gauge} for the embedding of five-dimensional $N=4$ gauged
$SU(2)\times U(1)$ supergravity in $D=10$.  The internal 5-sphere now
has a geometrical interpretation as a foliation by $S^3\times S^1$,
with $\xi$ parameterising the foliating surfaces.  

   Substituting 
(\ref{xscalar}), (\ref{gaugefields}) and (\ref{mus}) into the Ansatz
for $\hat G_\5$ given in (\ref{gans}), leads to
\bea
\hat{G}_\5 &=& -g\, U \,\varepsilon_5 - \frac{3 sc}{g} X^{-1}\,
{*dX}\wedge d\xi +
\frac{c^2}{8\sqrt{2}\, g^2} X^{-2}\, {*F^i_\2}\wedge h^j\wedge h^k
\, \varepsilon_{ijk} \nn\\
& &-\frac{s\,c}{2\sqrt{2}\,g^2} X^{-2}\, {*F^i_\2}\wedge h^i\wedge d\xi
- \frac{sc}{g^2} X^4\, {*G_\2}\wedge d\xi\wedge (d\tau - g B_\1)\,,
\eea
where $U= -2( X^2\, c^2 + X^{-1}\, s^2 + X^{-1})$, again in agreement
with the Ansatz obtained in \cite{d5gauge}.  As a consistency check,
we can also verify that substituting (\ref{xscalar}), 
(\ref{gaugefields}) and (\ref{mus}) into the Ansatz (\ref{gdualans})
for ${\hat *\hat G_\5}$ gives the same expression as the one obtained
for ${\hat *\hat G_\5}$ in \cite{d5gauge}.  Thus we have verified that
the gauged $SO(6)$ embedding that we have obtained in this paper
can be truncated to the $SU(2)\times U(1)$ embedding of the $N=4$
theory whose consistency was proven in \cite{d5gauge}.  This again
provides further supporting evidence for the consistency of the
our gauged $SO(6)$ reduction Ansatz.

\section{Consistency conditions for sphere reductions}

   One of the interesting outcomes from our analysis is that it is
essential for the consistency of the 5-form reduction Ansatz that it
be a {\it self-dual} 5-form, rather than an unconstrained one.  It
seems, therefore, that the requirement of consistency of the sphere
reduction has singled out a ten-dimensional starting point that is
itself embeddable in a supersymmetric theory.

   This raises the more general question of what possible
higher-dimensional theories might allow consistent Kaluza-Klein sphere
reductions.  All the known examples are associated with supersymmetric
higher-dimensional theories, but one might wonder whether this was just
a reflection of the fact that these are the cases that have received
the most attention in the literature.  However, the following argument
seems to suggest that the supersymmetric cases may be the only ones
that can allow consistent $S^n$ sphere reductions, in which all the
massless fields (including the $SO(n+1)$ Yang-Mills fields) are
retained.

   Consider a $D$-dimensional theory of gravity plus an $n$-form field
strength, with the Lagrangian
\be
e^{-1}\, {\cal L}_D = R - \ft1{2\, n!} \, F_n^2\,,\label{nlag}
\ee
where $e=\sqrt{-g}$.  If this were to give a $(D-n)$-dimensional
theory with an $SO(n+1)$ gauge group, as a consistent reduction on
$S^n$, it would be necessary that the ungauged $(D-n)$-dimensional
theory obtained by reducing instead on the torus $T^n$ should have a
global symmetry group $G$ that contains $SO(n+1)$ as a compact
subgroup, since an $SO(n+1)$ factor in the denominator group would be
gauged in the spherical reduction.  A reduction on $T^n$ always
produces a theory with a $GL(n,\R)$ global symmetry, which has $SO(n)$
as its maximal compact subgroup, and so this would be insufficient for
allowing an $SO(n+1)$ gauging.  In special cases the $GL(n,\R)$ global
symmetry can be enhanced \cite{cjlp13} to $SL(n+1,\R)$, but this
happens only if there is a ``conspiracy'' between axionic scalars
coming from the metric and axions coming from the form-field $F_n$.
For this conspiracy to occur, the strengths of the dilaton couplings
to axions from these two sources must be the same.\footnote{It is
always the case that the counting of dilatons and axions would be
consistent with the numerology required for an $SL(n+1,\R)/SO(n+1)$
coset structure, but this does not in general imply that the enhanced
coset actually occurs.}  Specifically, if $\vec\phi$ denotes the set
of $n$ canonically-normalised dilatons that result from the $T^n$
reduction then all the dilaton/axion couplings should be of the form
$e^{\vec a_i\cdot\vec\phi}\, (\del\chi_i)^2$, where the constant
vectors $\vec a_i$ satisfy $(\vec a_i)^2=4$ for each $i$.  (In fact
the full set of $\vec a_i$ vectors would constitute the positive-root
vectors of $SL(n+1,\R)$ \cite{cjlp13}.)  It was shown in
\cite{stainless} that the strengths of dilaton couplings can be
conveniently characterised in terms of the quantity $\Delta$, related
to the dilaton vector $\vec a$ by
\be
\vec a^2 = \Delta - \fft{2(n-1)(D-n-1)}{D-2}\,,\label{deltadef}
\ee
for an $n$-form field strength in $D$ dimensions, since the quantity
$\Delta$ is preserved under toroidal Kaluza-Klein reduction.
Furthermore, it was shown that the Kaluza-Klein vectors and axions
coming from a toroidal reduction of the metric always have $\Delta=4$.
It therefore follows that for the symmetry enhancement to
$SL(n+1,\R)$ to take place, the dilaton-coupling $\Delta$ for the
original $n$-form field\footnote{If there is no dilaton in the
original higher-dimensional theory then the ``dilaton coupling''
$\Delta$ is given by setting $\vec a^2=0$ in (\ref{deltadef}), and it is
this value that is preserved under toroidal reduction.}
in (\ref{nlag}) must also take the value $\Delta=4$.  An enumeration
of all possible cases shows that for Lagrangians of the form
(\ref{nlag}) one gets $\Delta=4$ only for 
\be
(D,n)\,=\, (11,4)\,,\qquad (11,7)\,,\qquad (10,5)\,.\label{three}
\ee

   The above considerations seem to single out the three cases in
(\ref{three}) as the only ones where an $S^n$ reduction of a
Lagrangian of the form (\ref{nlag}) could consistently yield the gauge
fields of $SO(n+1)$ and the scalars $T_{ij}$ of the coset
$SL(n+1,\R)/SO(n+1)$.\footnote{Consistent sphere reductions with
scalars $T_{ij}$ for more general $(D,n)$ values for the Lagrangian
(\ref{nlag}) have been suggested in \cite{nv}.  We expect that a more
complete analysis of the consistency of the reduction would exclude
such possibilities.  In particular, scalars $T_{ij}$ parametrising the
coset $SL(n+1,\R)/SO(n+1)$ only occur in the three cases listed in
(\ref{three}).} As we have seen in section 2 of this paper, it can
turn out that additional structure is also required in order to
achieve consistency, namely the self-duality of the 5-form in the case
of $(D,n)=(10,5)$.  By the same token one can expect that a consistent
reduction would only be possible in the cases $(D,n)=(11,4)$ and
$(11,7)$ if additional structure is also present in the
eleven-dimensional Lagrangian.

   For example, if we consider the $S^7$ reduction from $D=11$, then
the total set of 70 spin-0 fields decompose as a $35_v$ of scalars and
a $35_c$ of pseudoscalars.  The 35 scalars in $T_{ij}$ correspond to
the $35_v$.  Turning on these forces all the 28 gauge fields of
$SO(8)$ to be excited, and in turn these excite the remaining $35_c$
of pseudoscalars.  In order for an $S^7$ reduction that retains all
these fields to be consistent, it is necessary to include an
$F_\4\wedge F_\4\wedge A_\3$ term in the original Lagrangian
(\ref{nlag}), with precisely the coefficient dictated by $D=11$
supersymmetry.  This point also emphasises that in the $S^7$ reduction
one cannot consider just the 35 scalars $T_{ij}$ in isolation; the
full set of bosonic fields of $N=8$ gauged supergravity (and not
merely the 28 gauge fields) must be included if the full set of
$T_{ij}$ scalars are present.

    A similar situation arises with the $S^4$ reduction from $D=11$.
Including the full set of 14 scalars $T_{ij}$ in a consistent
reduction will force the complete set of massless fields to be
non-vanishing, including not only the ten Yang-Mills fields of $SO(5)$
but also the five 3-form field strengths coming from the antisymmetric
tensor.  The consistency of the reduction is then only possible if the
$FFA$ term of $D=11$ supergravity is included.

    Another possibility for obtaining further examples of consistent
sphere reductions is to include a dilaton in the higher-dimensional
Lagrangian, whose coupling to the $n$-form field strength is arranged
to have $\Delta=4$:
\be
e^{-1}\, {\cal L}_D = R -\ft12  (\del\phi)^2 -
\ft1{2\, n!} \, e^{a\, \phi}\, F_n^2\,,\label{nlag2}
\ee
with $a^2= 4- 2(n-1)(D-n-1)/(D-2)$.  (Lagrangians of this type without
the restriction on the value of the constant $a$ have also been
discussed in \cite{nv} in the context of sphere reductions.)  In this
case there will no longer be an AdS$_{D-n}\times S^n$ vacuum solution,
but rather a warped product of a domain wall and $S^n$.  The
possibilities for achieving $\Delta=4$ couplings are in fact rather
limited, given that $a$ should be a real number.  First of all, for
an $n$-form with $4\le n\le D-4$ it can be seen that we must have
$D\le 11$.  For example if $n=4$, then we must have
$D\le 11$, while for $n=5$ we must have $D\le 10$.  (It is only
necessary to consider forms with $n\le D/2$ since Hodge duality
maps those with $n>D/2$ into this range.)  Interestingly, for
$n=3$ one can achieve $\Delta=4$ in an arbitrary dimension $D$; the
Lagrangian (\ref{nlag2}) then corresponds to the low-energy effective
theory of the $D$-dimensional bosonic string.  We can expect that the
full 3-sphere reduction of this Lagrangian (keeping the complete
$SO(4)$ gauge fields of its isometry group, not merely the $SU(2)$
subset of left-invariant fields) will be consistent.  Similarly, we
can expect that it should be consistent to reduce the theory with
$n=3$ on a $(D-3)$-sphere.  Before the gauging, the scalar coset in
$D=3$ is $SO(D-2,D-2)/(SO(D-2)\times SO(D-2))$.  One of the $SO(D-2)$
denominator group factors can be gauged, and we obtain the scalar
coset $SL(D-2,\R)/SO(D-2)$ together with the additional gauge fields
of $SO(D-2)$, and a singlet scalar (which is the original dilaton of
the $D$-dimensional theory).

    A consistent sphere reduction, albeit of a slightly different
kind, has in fact been obtained in an example where there is a dilaton
in the higher-dimensional theory that couples to the form-field.  In
\cite{d6gauge} it was shown that a reduction of the massive type IIA
supergravity on an internal 4-dimensional space that is locally $S^4$
gives rise to the $N=2$ $SU(2)$ gauged supergravity in $D=6$.

\section{Conclusions and further comments}

    In this paper we have constructed a consistent Kaluza-Klein
reduction Ansatz for embedding the subset of the fields of
five-dimensional $N=8$ gauged supergravity, comprising gravity, the
fifteen $SO(6)$ gauge fields, and the twenty scalars of the
$SL(6,\R)/SO(6)$ submanifold of the full scalar manifold, into type
IIB supergravity in $D=10$.  This embedding can equivalently be viewed
as a complete reduction Ansatz (with no truncation of massless fields)
for the ten-dimensional theory comprising just gravity plus a self-dual
5-form, which itself is a consistent truncation of type IIB supergravity.

   A crucial point in the analysis is that in the gauged
five-dimensional supergravity one cannot consistently set the
Yang-Mills fields to zero, while retaining the full set of scalar
fields, unlike the situation in ungauged supergravity.  In the context
of the truncation that we consider in this paper, where we retain the
twenty scalars $T_{ij}$, we cannot ignore the fifteen $SO(6)$ gauge
fields, since the scalars act as sources for them.\footnote{This also
implies that solutions built using these scalars will in general
require non-vanishing Yang-Mills fields.}  We saw that the
self-duality of the 5-form field of the type IIB theory plays an
essential r\^ole in the consistency of the reduction.

   More generally, we showed that if one starts from a theory
comprising gravity and an $n$-form field strength, then a consistent
reduction on $S^n$ that retains the scalars $T_{ij}$ of the coset
$SL(n+1,\R)/SO(n+1)$ will also have to include at least the gauge
fields of $SO(n+1)$, and furthermore will only be possible for the
$S^4$ and $S^7$ reductions of $D=11$, and the $S^5$ reduction of
$D=10$.  In fact the consistency will in addition require that the
theories in $D=11$ and $D=10$ have the additional structures
associated with $D=11$ and type IIB supergravity, namely the $FFA$
term in $D=11$, and the self-duality of the 5-form in $D=10$.  In
$D=7$ the scalars and Yang-Mills fields must be supplemented by the
five 2-form potentials, while in $D=4$ they must be supplemented by
the 35 pseudo-scalars, in order to achieve consistency.  On the other
hand in $D=5$ no additional fields beyond the scalars $T_{ij}$ and the
gauge fields $A_\1^{ij}$ are required for consistency.  In fact in all
three cases this is related to the presence of the terms of the form
(\ref{ffterm}), bilinear in Yang-Mills fields.  In $D=7$ this term
acts as a source for the five 3-form fields; in $D=4$ it acts as a
source for the 35 pseudo-scalars; but in $D=5$ it acts as a ``source''
for the Yang-Mills fields themselves.  This special feature of the
$D=5$ gauged supergravity may have implications in the dual
four-dimensional $N=4$ super Yang-Mills theory.

   We also discussed the more general possibilities that might arise
if one includes a dilaton in the higher-dimensional theory.  The
possibilities for further examples of consistent sphere reductions
seem to be rather limited, as discussed in section 4.  

   In a full $S^5$ reduction of type IIB supergravity there will be
additional fields coming from the reduction of the NS-NS and R-R
2-form potentials, and from the dilaton and axion.  A complete
analysis of the $S^5$ reduction can therefore be expected to be
extremely complicated.  In particular, for example, the $10$ and
$\overline{10}$ of pseudo-scalars lead to a considerably more complicated
metric reduction Ansatz.  The Ansatz for a subset of the fields that
included one scalar and one pseudoscalar was derived in
\cite{d45gauge}, and in \cite{pw}.  However even in that case, the
construction of the Ansatz for the antisymmetric tensor fields is
rather involved, and has not yet been pushed to completion.

\section*{Acknowledgement}  C.N.P. is grateful to the University of
Pennsylvania for hospitality during the course of this work.

\end{document}